\documentclass[conference]{IEEEtran}
\IEEEoverridecommandlockouts
\usepackage{cite}
\usepackage{amsmath,amssymb,amsfonts}
\usepackage{algorithmic}
\usepackage{graphicx}
\usepackage{textcomp}
\usepackage{xcolor}
\usepackage{array, ragged2e, vcell, tabularray, diagbox}
\usepackage{subcaption, float}
\usepackage[hidelinks]{hyperref}
\def\BibTeX{{\rm B\kern-.05em{\sc i\kern-.025em b}\kern-.08em
    T\kern-.1667em\lower.7ex\hbox{E}\kern-.125emX}}
\begin{document}

\makeatletter
\newcommand{\linebreakand}{%
  \end{@IEEEauthorhalign}
  \hfill\mbox{}\par
  \mbox{}\hfill\begin{@IEEEauthorhalign}
}
\makeatother

\title{W2E (Workout to Earn): A Low Cost DApp \\based on ERC-20 and ERC-721 standards
\thanks{This work is the output of the ASEAN IVO \url{http://www.nict.go.jp/en/asean_ivo/index.html} project ``Agricultural IoT based on Edge Computing'' and financially supported by NICT \url{http://www.nict.go.jp/en/index.html}. \newline
    Corresponding author: Tran Thi Thuy Quynh (quynhttt@vnu.edu.vn).}
}

\author{
\IEEEauthorblockN{Do Hai Son\IEEEauthorrefmark{2}}
\IEEEauthorblockA{dohaison1998@vnu.edu.vn}
\and
\IEEEauthorblockN{Nguyen Danh Hao\IEEEauthorrefmark{3}}
\IEEEauthorblockA{haonguyen.uet@gmail.com}
\linebreakand
\IEEEauthorblockN{Tran Thi Thuy Quynh\IEEEauthorrefmark{3}}
\IEEEauthorblockA{quynhttt@vnu.edu.vn}
\and
\IEEEauthorblockN{Le Quang Minh\IEEEauthorrefmark{2}}
\IEEEauthorblockA{quangminh@vnu.edu.vn}
\linebreakand
\IEEEauthorrefmark{2} VNU Information Technology Institute, Hanoi, Vietnam \\
\IEEEauthorrefmark{3} VNU University of Engineering and Technology, Hanoi, Vietnam
}

\maketitle

\begin{abstract}
Decentralized applications (DApps) have gained prominence with the advent of blockchain technology, particularly Ethereum, providing trust, transparency, and traceability. However, challenges such as rising transaction costs and block confirmation delays hinder their widespread adoption. In this paper, we present our DApp named W2E - Workout to Earn, a mobile DApp incentivizing exercise through tokens and NFT awards. This application leverages the well-known ERC-20 and ERC-721 token standards of Ethereum. Additionally, we deploy W2E into various Ethereum-based networks, including Ethereum testnets, Layer 2 networks, and private networks, to survey gas efficiency and execution time. Our findings highlight the importance of network selection for DApp deployment, offering insights for developers and businesses seeking efficient blockchain solutions. This is because our experimental results are not only specific for W2E but also for other ERC-20 and ERC-721-based DApps.
\end{abstract}


\section{Introduction}\label{sec:intro}
DApps have been conceptualized long before blockchain technology. However, this term is only widely known along with the launch of Ethereum (Eth) blockchain network in 2014~\cite{Zheng20023}. DApp is a distributed application that operates on Peer-to-Peer (P2P) blockchain networks. Nowadays, the types of DApps are very diverse, e.g., finance, social, games, storage, and so on, supported on a website or mobile device. Since running on blockchain platforms, DApp inherits its security properties, e.g., trustworthiness, traceability, transparency, etc.

Nevertheless, there are a few challenges that lead to DApps not being as widely utilized as personal computer and mobile applications at present. The first challenge is the rapid increase in transaction costs, also known as gas in Ethereum~\cite{Liu2020}. To illustrate, the average transaction cost for a regular Ethereum native token (ETH) transfer is \$2.07 USD~\cite{Gas} on Jan. 2024. This number of ERC-20 standard-based transfer token functions is \$4.07 USD. The cost for an essential function such as transferring is too high compared to the fee-free banking transfer service. The second challenge is the delay due to block confirmation times. For example, Ethereum fixed the block confirmation times at $12$~seconds. That means, in the worst case, a transaction needs up to $12$~seconds to be confirmed and ready for further steps in DApps. Many blockchain protocols, consensus mechanisms, and smart contract (SM) standards are proposed to deal with these challenges~\cite{Belchior2021}. Based on the official statistics~\cite{Dapp}, DApps are deployed on a lot of blockchain networks. In~\cite{Samudaya2021}, the authors explored factors that should guide users' choices and ensure a seamless match between the DApp and the blockchain platform, such as simplicity, cost, performance, and so forth. However, there is currently no practical research available on the selection of an optimal network for DApps. Especially for small businesses, choosing an unsuitable deployment network can lead to unnecessary fees and time on the end users' side.
\begin{figure*}[!t]
    \centering
    \includegraphics[width=.7\linewidth]{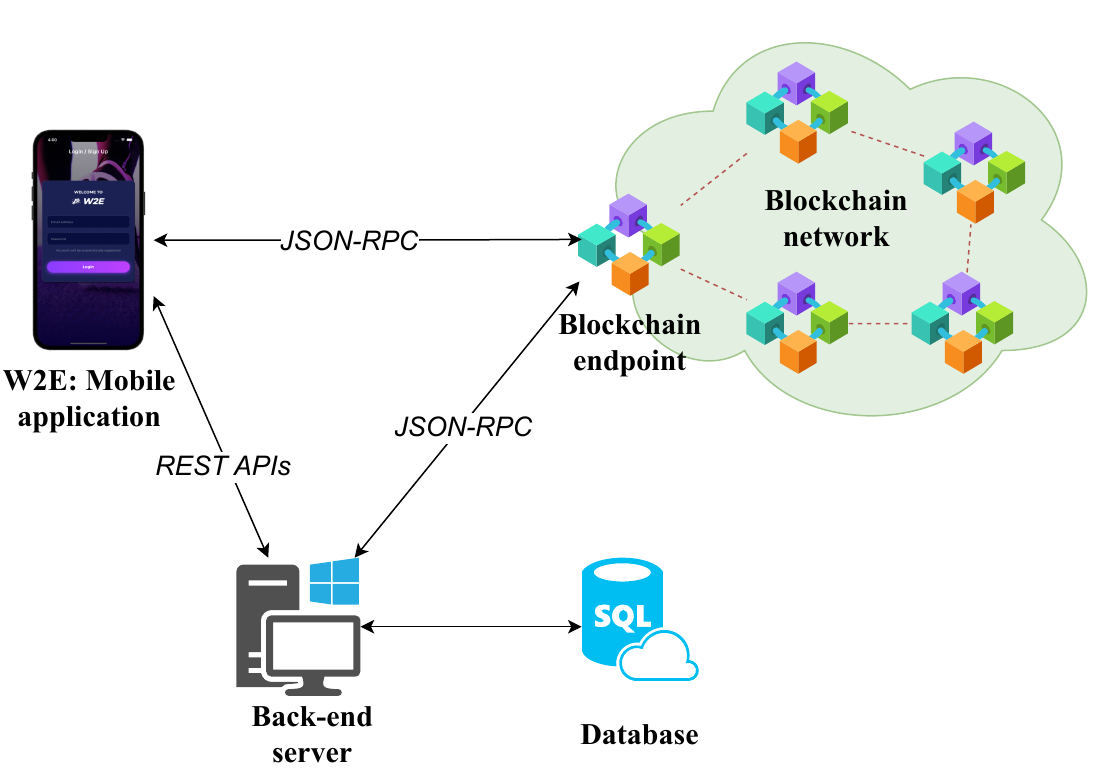}
    \caption{Architecture of W2E application.}
    \label{fig:system}
\end{figure*}

In this work, we focus on developing a mobile DApp called W2E - Workout to Earn. The objective of this application is to motivate users to exercise by rewarding them with Non-fungible tokens (NFTs) upon the completion of workout missions. W2E is a kind of general Web \& Contract as a DApp according to classification in~\cite{Zheng20023}. In detail, developers create a web or application front-end to link between smart contracts and typical missions, e.g., games, trading, or workouts, as in our project. The incentives are not directly real-world money but decentralized tokens, which can be exchanged for real-world money on trading markets, i.e., Binance. The tokens are generated through SMs, which are defined in Ethereum improvement proposals (EIPs). In this study, we use two popular token standards, i.e., ERC-20 and ERC-721. The former standard enables users to generate new tokens akin to ETH and subsequently conduct transfers, approvals, or minting. The latter standard represents ownership of NFT items, which can be images, items, and so forth. The link between ERC-20 and ERC-721 in W2E is that users do not directly buy ERC-721-based items by ETH but through our defined ERC-20 tokens, named DMD. DMD tokens can be purchased from the market by ETH or rewarded after completing workout missions. We then consider choosing an effective Ethereum-based blockchain network to deploy W2E DApp. Ethereum is now not only a standalone network but is the backbone of other decentralization solutions, i.e., SideChain (SC) and Layer 2 (L2)~\cite{Zhou2020}. Hence, we deploy our W2E on several Ethereum-based networks, i.e., Eth 1.0 testnet (Proof-of-Authority consensus mechanism - PoA), Eth 2.0 testnet (Proof-of-Stake consensus mechanism - PoS), Polygon testnet (L2), and Optimism testnet (L2). Moreover, besides testnet and mainnet on the public internet, a private Ethereum network is a good solution for specific purposes. For example, businesses can host a private Ethereum network and deploy W2E in this network, which has been proven to provide good efficiency in terms of gas and validation time~\cite{Son2021}. In this project, we built small-scale private Ethereum 1.0 and 2.0 networks~\cite{Son2021} to illustrate the private Ethereum solutions. As mentioned above~\cite{Samudaya2021}, we consider two important factors, i.e., gas consumption (money) and time consumption (delay), on these Ethereum-based blockchain networks. Our experiments reveal significant discrepancies in gas usage and operation time across various networks. Note that since W2E is developed based on well-known ERC-20 and ERC-721 standards, our experimental results also partly reflect practical DApps that use these two ERC standards.

The main contributions of this work are as follows: (i) the method to build W2E - a general DApp, its smart contracts source code is provided in~\footnote{\url{https://github.com/DoHaiSon/W2E_smart_contract}}; (ii) analysis of the gas and time consumption for several popular token standards on different Ethereum-based networks. The architecture and functions of W2E are presented in section~\ref{sec:w2e}. The deployment analysis of W2E's SM is shown in section~\ref{sec:ana}. Section~\ref{sec:concl} concludes this work and offers some potential research directions.

\section{Workout to Earn (W2E)}\label{sec:w2e}
\begin{figure*}[!t]
    \centering
    \includegraphics[width=.8\linewidth]{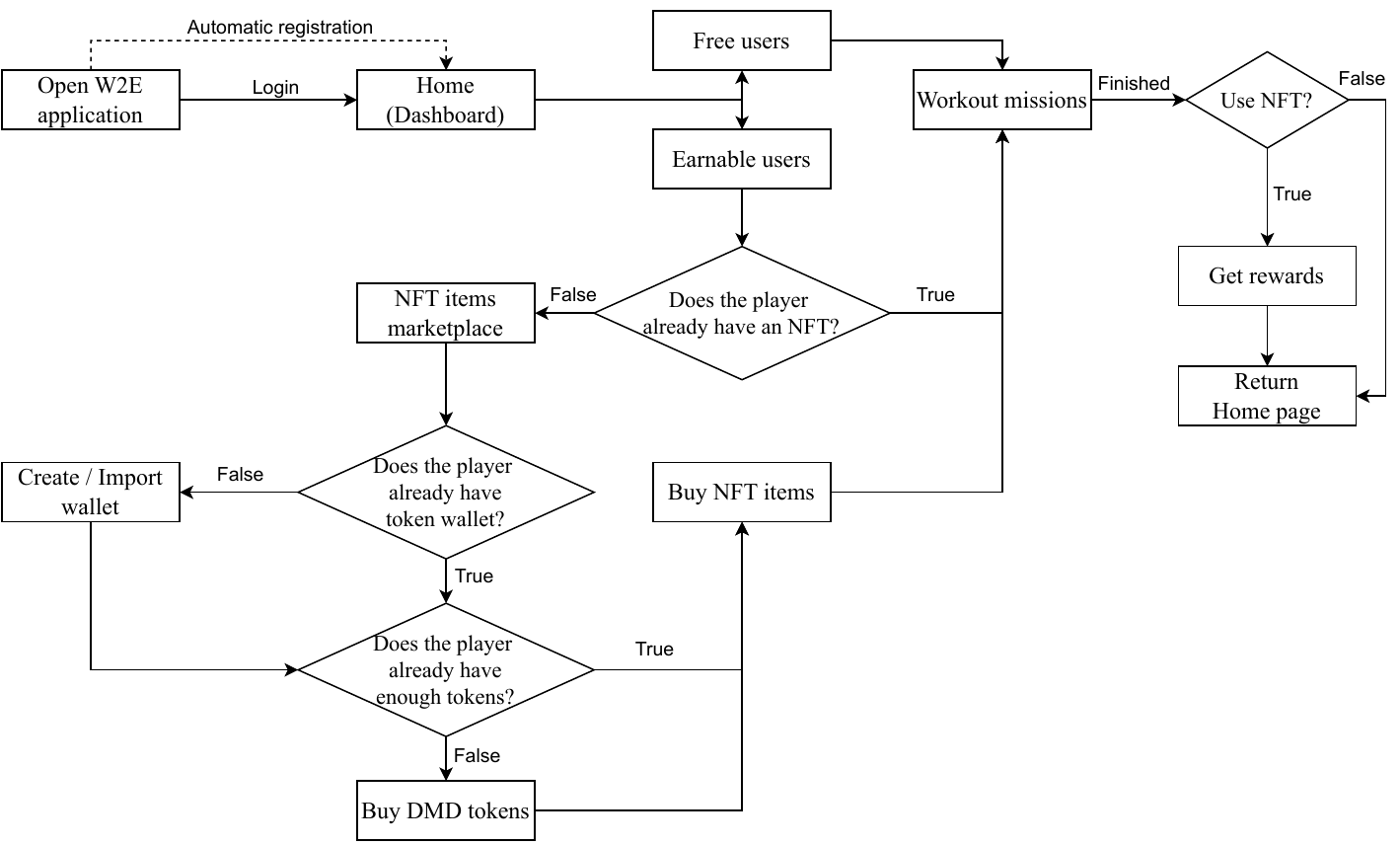}
    \caption{W2E: Users' flowchart.}
    \label{fig:flow}
\end{figure*}

W2E is a DApp that encourages exercise running on the iOS and Android platforms. W2E aims to reward users with a number of tokens (DMD) after finishing a workout, i.e., running. Users can sell their tokens (ERC-20) or use them to buy NFT items (ERC-721) on W2E's market. Both DMD tokens and NFT items (Pets) are tradeable. 

Fig.~\ref{fig:system} shows the architecture of the W2E application. There are three main components in W2E, as follows:
\begin{enumerate}
    \item W2E mobile application: this is the user interface of W2E where users can record their workout, earn DMD tokens, and trade NFT items. This component links to the back-end server to register and authenticate user accounts through the REST API protocol. On the other hand, the W2E application also connects to blockchain endpoints, i.e., Ethereum, Polygon, etc., through JSON-RPC connections. This link takes responsibility for sending transactions and loading users' balance information. W2E is developed on the Flutter framework.
    
    \item Blockchain endpoint: this is a node in the interested blockchain network, e.g., Ethereum mainnet/testnet/private networks, Solana, and so on. The node acts as a bridge to broadcast transactions of the W2E application and back-end server into the blockchain network via the synchronous process.
    
    \item Back-end server: this server accounts for several purposes, i.e., deploying smart contracts and storing users' account information and workout history in a centralized database. Additionally, the back-end server always listens to events from W2E smart contracts from the blockchain endpoint. That is a cache database for the W2E mobile application to load tokens and NFT information faster.
\end{enumerate}

Fig.~\ref{fig:flow} is the users' flowchart of W2E. Our application requires both an internet and GPS connection to track the users' velocity and distance. There are five main steps in W2E, as follows:
\begin{enumerate}
    \item Users log into their W2E account with email and password. W2E automatically registers an account if this device has never logged in before.

    \item Users import or create their external-owned account (EOA) into the W2E application. The creation phase is based on the mnemonic generation algorithm - BIP39~\cite{bip39}.

    \item Users require at least one pet to be able to receive rewards from workout missions. We call them ``Earnable users''. Each pet corresponds to a different reward rate for the user's training process. If users do not have any pets, they can buy them from the NFT items marketplace or act as non-profit users. Pets and DMD can be purchased with DMD tokens and ETH, respectively.

    \item Users engage in training missions. W2E captures metrics such as running time, distance, speed, and step count to help users track their workout progress.

    \item At the end, W2E rewards users an amount of DMD tokens based on their workout record and the bonus rate for the selected pet.
\end{enumerate}
\begin{figure*}
     \centering
     \begin{subfigure}[b]{0.48\textwidth}
         \centering
         \includegraphics[width=\textwidth]{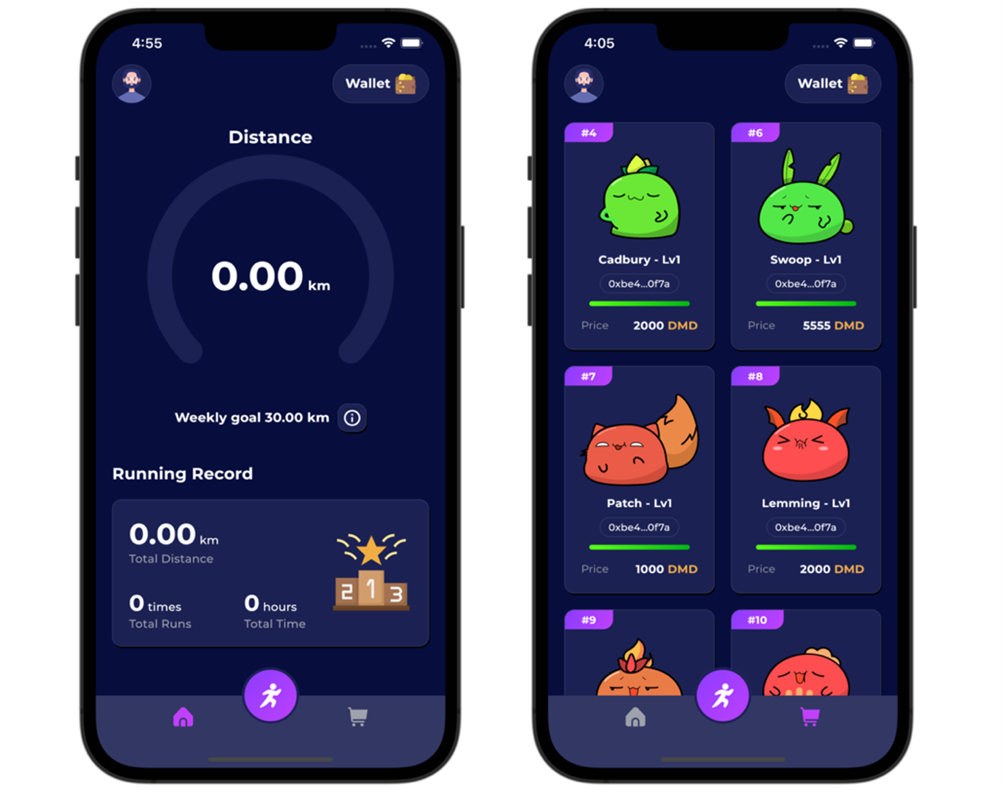}
         \caption{Dashboard and Market}
         \label{fig:a1}
     \end{subfigure}
     \hfill
     \begin{subfigure}[b]{0.48\textwidth}
         \centering
         \includegraphics[width=\textwidth]{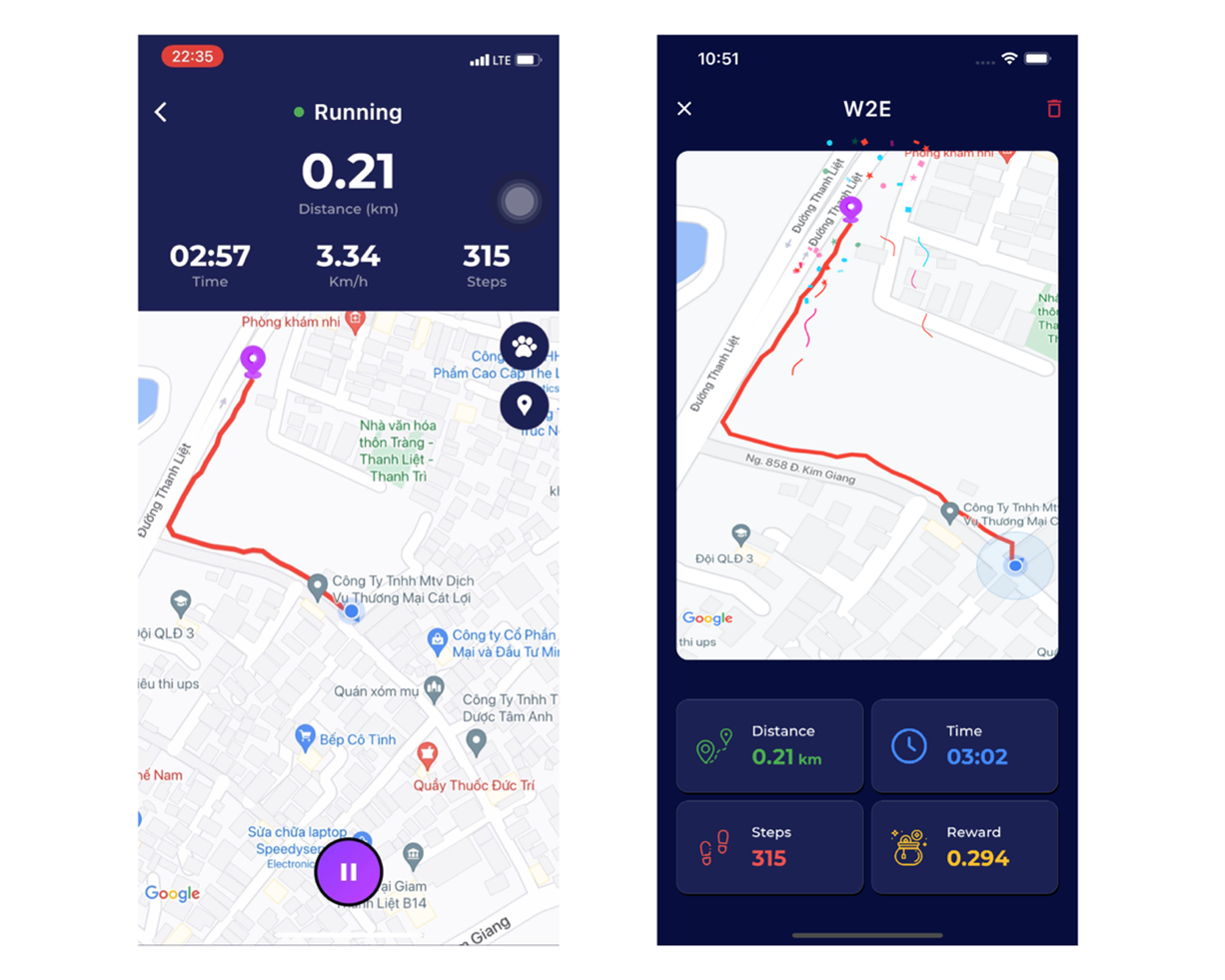}
         \caption{Workout record and rewards}
         \label{fig:a2}
     \end{subfigure}
     \hfill
     \begin{subfigure}[b]{0.48\textwidth}
         \centering
         \includegraphics[width=\textwidth]{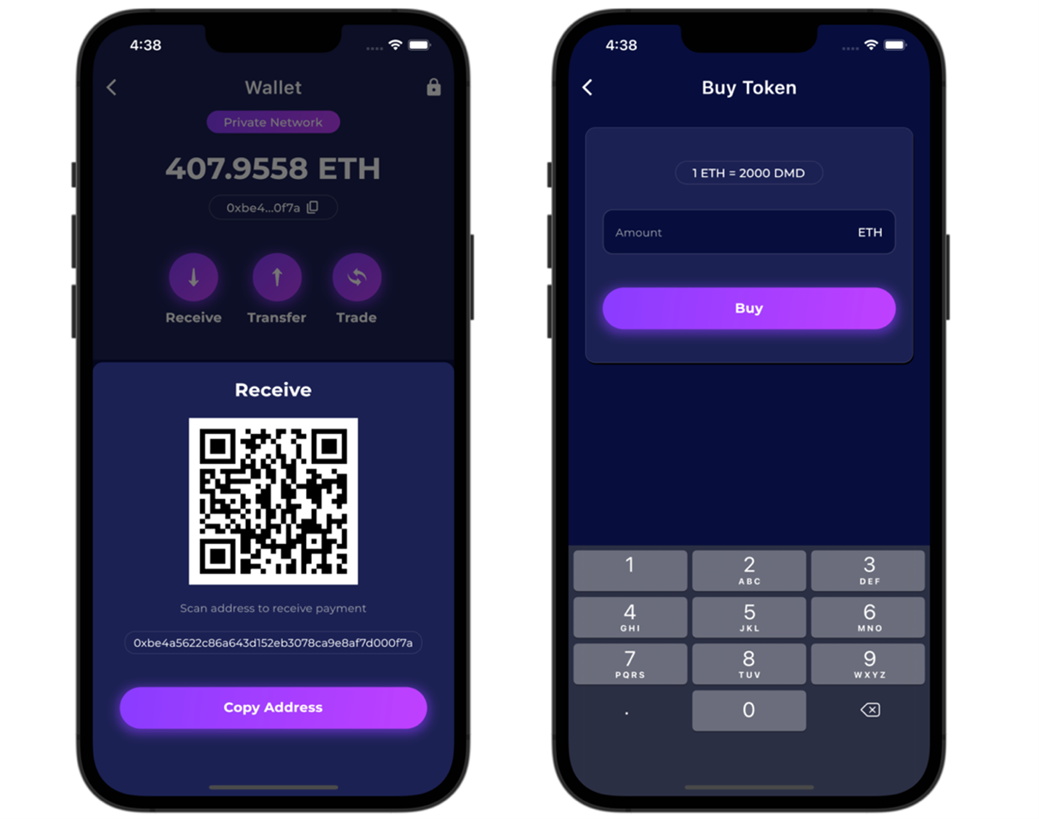}
         \caption{Buy DMD tokens (ERC-20)}
         \label{fig:a3}
     \end{subfigure}
     \hfill
     \begin{subfigure}[b]{0.48\textwidth}
         \centering
         \includegraphics[width=\textwidth]{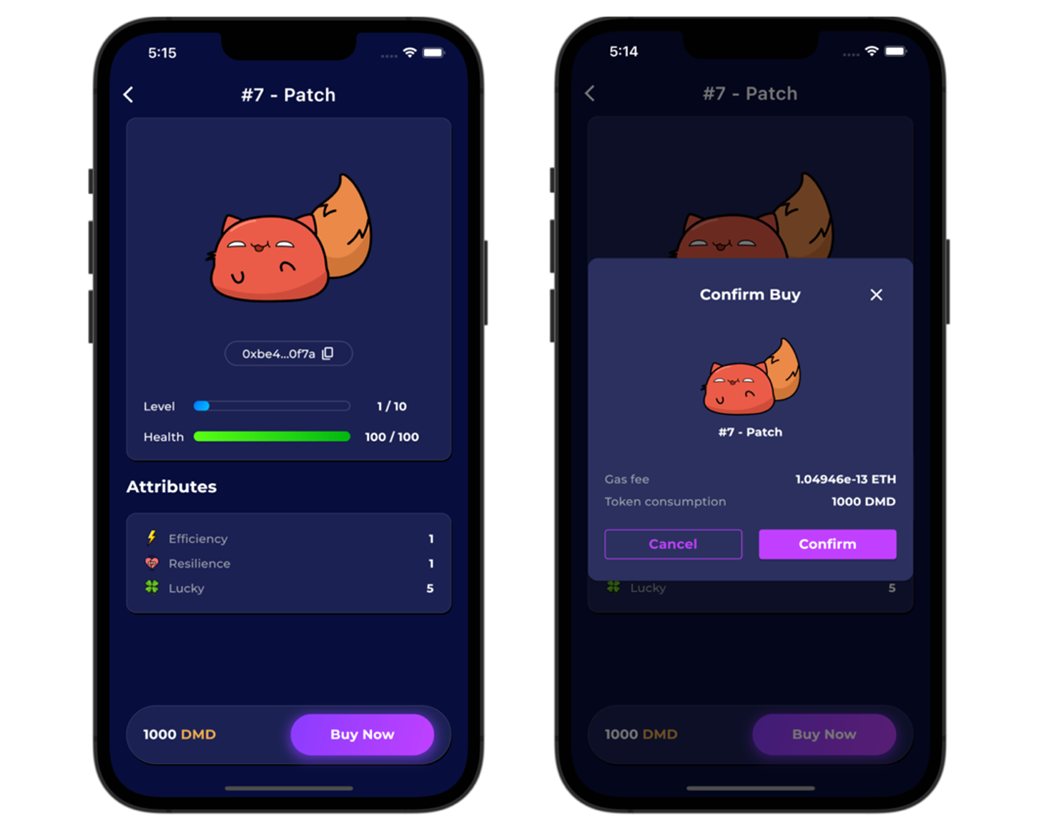}
         \caption{Buy an NFT item (ERC-721)}
         \label{fig:a4}
     \end{subfigure}
    \caption{Interfaces of main functions on W2E application.}
    \label{fig:w2e}
\end{figure*}
In detail, smart contracts of ERC-20-based DMD tokens and ERC-721-based NFT items are provided in our public repository:~\url{github.com/DoHaiSon/W2E_smart_contract}. There are two approaches to determining the price of DMD tokens,  which serve as monetary rewards for users. First, the business that released the W2E application sets the value of DMD tokens, using them as incentives for their users. Alternatively, DMD's price can be influenced by market dynamics, reflecting the demand and supply generated by players who trade this token, a common approach adopted by various DApp gaming applications. Fig.~\ref{fig:w2e} is some real screenshots captured from the W2E application. The dashboard and NFT marketplace interfaces are shown in Fig.~\ref{fig:a1}. Then, Fig.~\ref{fig:a2} is a record of a running mission on the W2E application. The interfaces of buying DMD tokens and acquiring pets are illustrated in Fig.~\ref{fig:a3} and Fig.~\ref{fig:a4}, respectively.

\section{Low Cost Deployment Analysis}\label{sec:ana}
\begin{figure}[H]
    \centering
    \includegraphics[width=\linewidth]{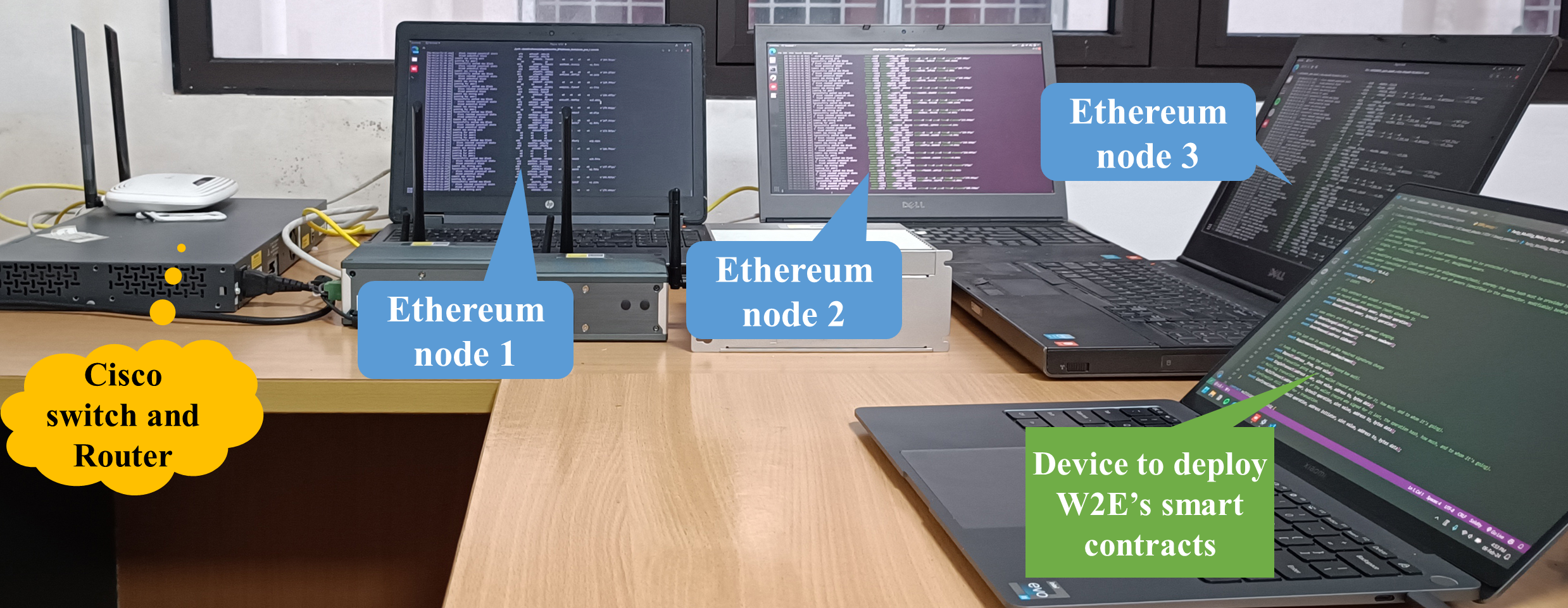}
    \caption{A small-scale private Ethereum 1.0 and 2.0 network configuration.}
    \label{fig:real}
\end{figure}

In this section, we deploy W2E smart contracts on several Ethereum-based networks to evaluate performance discrepancies across various implementations. Table~\ref{tab:network} presents a list of six networks that are considered in this work. Four of these networks are public Ethereum-based blockchain testnets available on the internet. Their configurations closely resemble the mainnet, but their tokens are non-valuable. We do not set up these networks but use APIs of~\url{infura.io} as a bridge between the W2E application and the blockchain endpoint. The remaining two networks are private, managed by us, and configured differently from both the mainnet and testnet. They are set up in our laboratory, as shown in Fig.~\ref{fig:real}. This small-scale setup has three Ethereum nodes and a device for developing the W2E application. All devices are connected to a local network by a Cisco switch. However, any business can adopt and host a blockchain network. The private Eth 1.0 network is launched using an older version of Ethereum and modified to reduce the block confirmation times, according to our previous work~\cite{Son2021}.
\begin{table*}
\centering
\caption{The list of deployment networks.}
\label{tab:network}
\begin{tblr}{
  width = .7\linewidth,
  colspec = {Q[294]Q[363]Q[237]Q[108]},
  row{1} = {c},
  cell{2}{4} = {c},
  cell{3}{4} = {c},
  cell{6}{4} = {c},
  cell{7}{4} = {c},
  hlines,
  vlines,
}
\textbf{Name} & \textbf{Network config} & \textbf{Source code / API} & \textbf{Version}\\
\textbf{Eth 1.0 testnet} & Goerli testnet & Infura.io~\cite{infura} & -\\
\textbf{Eth 2.0 testnet} & Ethereum-Sepolia testnet & Infura.io~\cite{infura} & -\\
\textbf{Private Eth 1.0} & Custom config~\cite{Son2021} & go-ethereum~\cite{Go-Ethereum} & 1.10.4\\
\textbf{Private Eth 2.0} & Mainnet config & {go-ethereum~\cite{Go-Ethereum}\\prysm~\cite{Prysm}} & {1.10.22\\3.2.0}\\
\textbf{Polygon testnet} & Mumbai testnet & Infura.io~\cite{infura} & -\\
\textbf{Optimism testnet} & Optimism-Sepolia testnet & Infura.io~\cite{infura} & -
\end{tblr}
\end{table*}
\begin{table*}
\centering
\caption{Gas consumption to deploy the W2E and other ERC-based smart contracts in Gwei.}
\label{tab:deployment}
\begin{tblr}{
  width = .7\linewidth,
  colspec = {Q[294]Q[160]Q[158]Q[160]Q[160]},
  column{even} = {r},
  column{3} = {r},
  column{5} = {r},
  cell{1}{2} = {c,t},
  cell{1}{3} = {c,t},
  cell{1}{4} = {c,t},
  cell{1}{5} = {c,t},
  hlines,
  vlines,
}
\diagbox[width=\dimexpr\linewidth+2\tabcolsep, height=25pt]{\textbf{Network}}{\textbf{Token}} & {\textbf{ERC-20}\\(W2E)} & {\textbf{ERC-721}\\(W2E)} & \textbf{ERC-1155} & \textbf{ERC-777}\\
\textbf{Eth 1.0 testnet} & 5644679.62 & 5611030.26 & 5543731.55 & 5602617.92\\
\textbf{Eth 2.0 testnet} & 395379.94 & 378555.26 & 353318.25 & 370142.93\\
\textbf{Private Eth 1.0} & 6.84 & 6.35 & 6.44 & 6.58\\
\textbf{Private Eth 2.0} & 0.05 & 0.04 & 0.04 & 0.04\\
\textbf{Polygon testnet} & 48791.57 & 50474.04 & 41220.46 & 42902.93\\
\textbf{Optimism testnet} & 82440.92 & 92535.73 & 79917.22 & 117772.75
\end{tblr}
\end{table*}

We first examine the gas consumption of a few ERC-based smart contracts, i.e., ERC-20, ERC-721, ERC-1155, and ERC-777. As shown in Table~\ref{tab:deployment}, the fees for deploying W2E's and other SMs on networks exhibit significant discrepancies. When considering the testnets, it is evident that two L2 networks significantly reduce gas consumption. L2 networks were specifically designed to lower transaction costs, which sets them apart from Ethereum 1.0 and 2.0 networks. Particularly noteworthy is the Polygon testnet, which requires only about 10\% of the gas necessary to deploy SMs compared to the Eth 2.0 testnet. Eth 1.0 testnet horribly consumes gas, which led to the Eth 2.0 upgrade. In contrast, private networks, owing to custom configurations, entail trivial deployment fees. Specifically, deploying ERC-721, ERC-1155, and ERC-777 contracts incurs fees of approximately $0.04$ Gwei.
\begin{table*}[]
\centering
\caption{Gas consumption to execute W2E's functions in Gwei.}
\label{tab:funct}
\begin{tblr}{
  width = .7\linewidth,
  colspec = {Q[312]Q[196]Q[202]Q[210]},
  column{even} = {r},
  column{3} = {r},
  cell{1}{2} = {c},
  cell{1}{3} = {c},
  cell{1}{4} = {c},
  hlines,
  vlines,
}
\diagbox[width=\dimexpr\linewidth+2\tabcolsep]{\textbf{Network}}{\textbf{Function}} & \textbf{Buy NFT} & \textbf{Sell NFT} & \textbf{Cancel NFT}\\
\textbf{Eth 1.0 testnet} & 6511150.56 & 5737215.35 & 4130458.56\\
\textbf{Eth 2.0 testnet} & 529977.37 & 429029.3 & 319668.89\\
\textbf{Private Eth 1.0} & 10.85 & 7.95 & 4.78\\
\textbf{Private Eth 2.0} & 0.09 & 0.07 & 0.03\\
\textbf{Polygon testnet} & 74028.59 & 59727.61 & 43744.16\\
\textbf{Optimism testnet} & 126185.09 & 92535.73 & 74869.82
\end{tblr}
\vspace*{-0.3cm}
\end{table*}

While the deployment fee for SMs may be substantial, it is a one-time expense. Therefore, our primary concern lies in the transaction costs associated with activities, such as purchasing and selling NFTs or DMD tokens. Table~\ref{tab:funct} shows the execution fees for W2E's smart contract functions, i.e., buy, sell, and cancel NFT items. The results are close to deployment fees where Polygon testnet and private Eth 2.0 outperform other networks. We can conclude that if businesses want to deploy their applications onto Ethereum-based networks, they should consider SC or L2 solutions, e.g., Polygon or Optimism, to reduce gas consumption. On the other hand, private network solutions may fit specific purposes, e.g., small businesses want to control their blockchain network themselves and do not want users to bear transaction fees.

We then investigate time consumption/delay due to transaction confirmation time. This directly impacts the user experience, as simple operations such as buying tokens can take a lot of time. Table~\ref{tab:deploytime} is the deployment time of the DMD token at five different timestamps. Initially, there was a difference in smart contract deployment time at different timestamps, which lasted up to over $40$~seconds. This is understandable because the deployment transaction can come at a time when the block is almost to be validated (good case) or take a long time for the block to be validated (bad case). Considering different networks, the private Eth 2.0 still stands out as the best solution, with a delay of almost less than $4$~seconds. In contrast to gas consumption, the Optimism testnet outperforms the Polygon and two Eth testnets. This performance variation reflects a trade-off between gas consumption and time delay considerations of the Polygon and Optimism networks.

Finally, we consider the time consumption to execute the buy NFT transactions of the W2E application at five timestamps. The experimental results in Table~\ref{tab:funttime} show similarities with Table~\ref{tab:deploytime}. If testnets are used, W2E publishers consider using L2 networks to reduce transaction latency. On the contrary, although it reduces decentralization, private Eth 2.0 networks provide significant efficiency compared to testnet solutions.
\begin{table*}[!t]
\centering
\caption{Time consumption to deploy the DMD token SM at different timestamps.}
\label{tab:deploytime}
\begin{tblr}{
  width = .7\linewidth,
  colspec = {Q[200]Q[65]Q[65]Q[65]Q[65]Q[65]Q[79]},
  column{even} = {r},
  column{3} = {r},
  column{5} = {r},
  column{7} = {r},
  cell{1}{2} = {c},
  cell{1}{3} = {c},
  cell{1}{4} = {c},
  cell{1}{5} = {c},
  cell{1}{6} = {c},
  cell{1}{7} = {c},
  hlines,
  vlines,
}
\diagbox[width=\dimexpr\linewidth+2\tabcolsep]{\textbf{Network}}{\textbf{Timestamp}} & {\textbf{1st}\\(ms)} & {\textbf{2nd}\\(ms)} & {\textbf{3rd}\\(ms)} & {\textbf{4th}\\(ms)} & {\textbf{5th}\\(ms)} & {\textbf{Average}\\(ms)}\\
\textbf{Eth 1.0 testnet} & 38489 & 38887 & 36808 & 40094 & 35500 & 37955.6\\
\textbf{Eth 2.0 testnet} & 20484 & 20491 & 21660 & 18660 & 17578 & 19774.6\\
\textbf{Private Eth 1.0} & 7057 & 5104 & 5758 & 5980 & 6830 & 6145.8\\
\textbf{Private Eth 2.0} & 3849 & 3849 & 3775 & 3539 & 4104 & 3823.2\\
\textbf{Polygon testnet} & 10788 & 10484 & 9771 & 8074 & 10920 & 10007.4\\
\textbf{Optimism testnet} & 5880 & 6196 & 5021 & 4275 & 5024 & 5279.2
\end{tblr}
\end{table*}
\begin{table*}[!htb]
\centering
\caption{Time consumption to execute the W2E's buy NFT function at different timestamps.}
\label{tab:funttime}
\begin{tblr}{
  width = .7\linewidth,
  colspec = {Q[200]Q[65]Q[65]Q[65]Q[65]Q[65]Q[79]},
  row{1} = {c},
  cell{2}{2} = {r},
  cell{2}{3} = {r},
  cell{2}{4} = {r},
  cell{2}{5} = {r},
  cell{2}{6} = {r},
  cell{2}{7} = {r},
  cell{3}{2} = {r},
  cell{3}{3} = {r},
  cell{3}{4} = {r},
  cell{3}{5} = {r},
  cell{3}{6} = {r},
  cell{3}{7} = {r},
  cell{4}{2} = {r},
  cell{4}{3} = {r},
  cell{4}{4} = {r},
  cell{4}{5} = {r},
  cell{4}{6} = {r},
  cell{4}{7} = {r},
  cell{5}{2} = {r},
  cell{5}{3} = {r},
  cell{5}{4} = {r},
  cell{5}{5} = {r},
  cell{5}{6} = {r},
  cell{5}{7} = {r},
  cell{6}{2} = {r},
  cell{6}{3} = {r},
  cell{6}{4} = {r},
  cell{6}{5} = {r},
  cell{6}{6} = {r},
  cell{6}{7} = {r},
  cell{7}{2} = {r},
  cell{7}{3} = {r},
  cell{7}{4} = {r},
  cell{7}{5} = {r},
  cell{7}{6} = {r},
  cell{7}{7} = {r},
  hlines,
  vlines,
}
\diagbox[width=\dimexpr\linewidth+2\tabcolsep]{\textbf{Network}}{\textbf{Timestamp}} & {\textbf{1st}\\(ms)} & {\textbf{2nd}\\(ms)} & {\textbf{3rd}\\(ms)} & {\textbf{4th}\\(ms)} & {\textbf{5th}\\(ms)} & {\textbf{Average}\\(ms)}\\
\textbf{Eth 1.0 testnet} & 39492 & 38004 & 38808 & 42560 & 41224 & 40017.6\\
\textbf{Eth 2.0 testnet} & 18600 & 18112 & 19750 & 21320 & 19682 & 19492.8\\
\textbf{Private Eth 1.0} & 6168 & 6100 & 4867 & 6480 & 6941 & 6111.2\\
\textbf{Private Eth 2.0} & 4250 & 4200 & 3744 & 3899 & 4007 & 4020\\
\textbf{Polygon testnet} & 11000 & 12982 & 10026 & 10145 & 9860 & 10802.6\\
\textbf{Optimism testnet} & 5120 & 6087 & 5000 & 4384 & 4932 & 5104.6
\end{tblr}
\end{table*}

\section{Conclusion}\label{sec:concl}

This study introduced the W2E application and investigated the implementation costs across various blockchain solutions. We first described the architecture and users' flowchart of the W2E application. We then deployed W2E's smart contracts onto six Ethereum-based networks. Through experiments, L2 solutions or private networks offer enhanced efficiency in gas usage and transaction execution time. Since W2E is a general DApp, developers can adopt the idea for their project. In the future, we will consider the performance of a DApp when the number of blockchain nodes is scaled up.

%
\bibliographystyle{IEEEtran}
\bibliography{library}
\end{document}